# The Weak Blue Bump of H2106-099

# and

# AGN De–Reddening

**B. Grossan**[1]
bruce@singu.lbl.gov

**R. A. Remillard**
rr@space.mit.edu

**H. V. Bradt**
hale@space.mit.edu

Center for Space Research, 37-578, Massachusetts Institute of Technology, Cambridge, MA 02139

**R. J. V. Brissenden**
Harvard-Smithsonian Center for Astrophysics, 60 Garden St., Cambridge, MA 02138
rjb@cfa.harvard.edu

**T. Ohashi**
Department of Physcis, Tokyo Metropolitan University,1-1 Minami-Ohsawa, Hachioji, 192-03, Japan, ohashi@phys.metro-u.ac.jp

**T. Sakao**
National Astronomical Observatory, Mitaka, Tokyo 181, Japan,
sakao@spot.mtk.nao.ac.jp

[1]Currently at University of California Space Science Laboratory, Postal Address: Lawrence Berkeley Laboratory, M/S 50-232, 1 Cyclotron Road, Berkeley, CA 94720

# ABSTRACT

We present multi-frequency spectra of the Seyfert 1 galaxy H2106-099, from radio to hard X-rays, spanning over a decade of observations. The hard X-ray (2-20 keV) spectrum measured with Ginga was not unusual, with a Log slope ($F_\nu \propto E^\alpha$), $\alpha_X$, of –0.80 ± 0.02 on 1988 May 18 and –1.02 ± 0.10 on 1988 May 22 / 23 UT, and with no significant observed variations in total flux. Other measurements showed variability and unusual spectral features: The V band flux was observed to change by a factor of 1.8 (>10 $\sigma$) in six weeks. Only moderate optical Fe II emission is present, but strong [FeVII] and [Fe X] lines were present in some epochs. The Balmer lines show > 25% variations in flux relative to the mean, and He I changed by more than 100% relative to the mean in <~ six years. The most surprising finds came from the composite UV through near-IR spectrum: If the spectrum is de-reddened by the galactic extinction value derived from 21 cm observations, a residual 2175 Å absorption feature is present. Additional de-reddening to correct the feature yields *E(B–V)*=0.07 mag due to material *outside our galaxy,* most probably associated with the AGN or its host galaxy. No other clear indications of reddening are observed in this object, suggesting that blue bump strength measurements in low and intermediate red-shift AGN could be incorrect if derived without UV observations of the region near 2175 Å in the AGN frame. After all reddening corrections are performed, the log slope of H2106-099 from the near IR (~12500 Å) to the UV (~1400 Å), –0.94 ± 0.05, is steep compared to other AGN, suggesting that the blue bump in this object is *intrinsically* weak. Weak blue bumps are, therefore, not always an artifact caused by reddening. The spectral indices of this object in the optical-UV region are steeper than those predicted by optically thin free-free emission models; therefore, some other mechanism must dominate the continuum in this region.

## 1. INTRODUCTION

AGN, or Active Galactic Nuclei, are powerful extragalactic objects whose nuclear emissions cannot be explained as products of normal stellar evolution. In this paper, the term AGN will be taken to mean emission-line AGN without significant optical polarization (non-blazar AGN). These nuclei are often more luminous than the whole of typical galaxies, emitting roughly equal power per decade over most observable frequencies higher than the radio bands.

The basic features of typical AGN spectra can be described by a combination of a radio component, a small IR bump (thermal emission from dust), an IR-UV continuum (often modeled as a power law), emission lines, a "blue bump" (possibly thermal emission from an accretion disk), and an X-ray power law, typically with a Log slope ($F_\nu \sim E^\alpha$) of $\alpha_X \sim -0.7$ (Turner & Pounds 1989). Energy and variability time scale arguments suggest that AGN are powered by accretion onto a super-massive black hole ($10^6$ - $10^9$ $M_{solar;}$ e.g. Blanford 1990). The accreting matter is believed to be the source of X-ray continuum radiation which we observe directly, but the emission mechanism is not understood and reprocessed radiation may dominate the spectrum at higher X-ray energies than are discussed here (Guilbert & Rees 1988). The X-ray / UV continuum photoionizes material at radii much larger than that of the X-ray emitting region to produce the broad emission lines observed in optical and UV spectra.

Deconvolution of the various spectral components can be difficult, particularly in the UV-optical region, because of the presence of blended multiplets of Fe II emission lines forming an apparent continuum over much of the region from 2000 to 5000 Å. Extensive modeling of quasar spectra has been done by Wills, Netzer, and Wills (1985, hereafter “WNW”), and of Seyfert galaxy spectra by Netzer et al. (1985), to understand the Fe II emission, allowing separation of this component of the spectrum. Below, we make use of this work in spectral fitting to interpret our multi-frequency observations of H2106-099.

The presence of an accretion disk is a natural consideration for any accreting system in order to conserve angular momentum. The optical-UV blue bump in AGN can be modeled by a combination of black body spectra (e.g. Malkan 1983), and is generally associated with the emission from an optically thick accretion disk. The confirmation of Seyfert galaxies with no or extremely weak blue bumps might suggest alternative AGN accretion models. McDowell et al. (1989) found that a number of quasars have little or no blue bump. However, the apparent weakness of these bumps could be due to insufficient reddening corrections.

Extragalactic observations may be corrected for reddening within our galaxy by making use of the correlation between $N_{HI}$ , the neutral hydrogen column density, and the dust absorbing column (Van Steenberg & Shull 1988). This correlation has a particularly small scatter for high-latitude extragalactic observations, as any small scale clumping is averaged out over the longer lines of sight through the full height of the disk (Savage & Mathis 1979). The column density of neutral hydrogen, $N_{HI}$ , within the Milky Way can be obtained from the results of 21 cm emission surveys. The typical correction for reddening in the UV, near Ly$\alpha$, is greater than a factor of two, so large uncertainties in the reddening correction can significantly affect the measured strength of a blue bump. Perhaps the greatest uncertainty in reddening corrections for AGN spectra comes from the unknown extinction associated with the host galaxy or the AGN itself. Grandi (1983) gives a thorough discussion of the various indicators of reddening, but concludes that there are problems with every type of indicator. A prominent dust absorption feature at 2175 Å exists, and seems like an obvious candidate for an indicator . Grandi (1983) criticized the use of this feature as a reddening indicator for the most part because it had been observed only rarely, although a number of AGN appear to be reddened when evaluated with other indicators. Since the aforementioned article, however, several obvious, deep 2175 Å features have been observed in AGN (e.g. Kinney et al. 1991). In this paper we demonstrate the use of the 2175 Å feature to measure reddening and give a blue bump strength measurement corrected for this effect.

The Large Area Sky Survey (LASS; Wood et al. 1984) source 1H2107–097 has been identified with a Seyfert Type 1 galaxy with z=0.027 designated H2106–099 (Remillard et al. 1986). The weighted mean X-ray luminosity of this source, as determined by our 1988 May measurements (described below), is $4.7 \times 10^{43}$ erg $s^{-1}$ ($H_0$/50 km $s^{-1}$ $Mpc^{-1})^{-2}$ in the 2-10 keV band. The optical spectrum of the object shows moderate Fe II strength and high excitation emission lines including [FeVII]λ5721, [FeVII]λ6086 and [Fe X]λ6374. The LASS instrument recorded a flux of $2.25 \pm 0.39 \times 10^{-11}$ ergs $s^{-1}$ $cm^{-2}$ (2–10 keV) during 1977 November 5–11, and a flux of $0.75 \pm 0.47 \times 10^{-11}$ ergs $s^{-1}$ $cm^{-2}$ (2–10 keV) during 1978 May 3–9, suggesting that the object could be variable. As part of a program to characterize the HEAO–1 discovered objects, we performed comprehensive, nearly-simultaneous, spectral observations of this object during or near 1988 May. The results of these observations are given below.

## 2. OBSERVATIONS AND ANALYSIS

We observed H2106-099 with the Ginga and IUE satellites and the McGraw-Hill 1.3m telescope at MDM* during 1988 May. Portions of these observations were simultaneous, while radio measurements with the Very Large Array (VLA) were acquired two weeks earlier, and Near IR observations had been performed 11 months earlier. The object had been observed before the Ginga observations in the X-ray, UV, IR, and radio

* Observations reported herein were carried out, in part, at the Michigan-Dartmouth-M.I.T. (MDM) observatory, Kitt Peak, which is operated by a consortium consisting of the University of Michigan, Dartmouth College, and M.I.T.

bands, and has been monitored before and after the Ginga observation in the optical band. A journal of observations is given in Table 1.

All photometric measurements at the MDM observatory after 1988 were acquired with Johnson system filters using the Michigan Direct Camera and the Thomson CCD detector on the McGraw-Hill 1.3 m telescope. These data were reduced with the program VISTA (written by Todd Lauer). Fluxes were measured in 15" square apertures. Photometry during 1987 May was measured with a photoelectric detector with a 10” diameter circular aperture, described in Brissenden (1989), and photometry during 1982 is described in Remillard et al. (1986). Observations of photometric standards from Landolt (1973, 1983) were performed during all observing runs.

The MDM optical spectra during 1988 May were obtained with the 1.3 m telescope using the Mark III spectrograph, with the same CCD detector as above, using a 5.2" slit. Spectrophotometry was normalized using measurements of standard stars of Oke (1974) or Stone (1977). Spectra taken in 1988 had 12 Å resolution, spectra taken in 1989 May and October had 6 Å resolution, with 4.3” and 3.3” slits, respectively. The spectra were reduced using the IRAF software package, described in Tody (1986).

The reduction of the IUE data occurred at the Regional Data Analysis Facility at Goddard Spaceflight Center, using the standard techniques and calibrations provided for IUE guest observers. The VLA data were taken in configuration "C" and analyzed with standard software provided by the VLA. The reduction of previously published data is described in the references to Table 1.

H2106–099 was observed with the LAC (Large Area Counter) instrument on the satellite Ginga on May 18 and again on May 22/23 of 1988. The LAC instrument is sensitive to X-rays from 2-50 keV and is described in Turner et al. (1989). The satellite was operated in the low bit rate MPC-1 mode, yielding 16 sec. time resolution. The results presented here were background subtracted using Method II, as described in Hayashida et. al. (1989), which models the background for subtraction in bins of 64 seconds or larger.

The background was also subtracted by Method I, which used off-source background observations made on May 17. There was no significant difference between the results from the two methods. The background subtracted spectra were fit to thermal bremsstrahlung, blackbody, and power law spectra, all with absorption by cold material. A power law yielded the best fit. The fits were made with software provided by the Ginga team, which minimizes the $\chi 2$ value of the fits. Gaussian emission profiles were added to the power law model to fit a small-scale residual bump near 6 keV. Such a bump could be caused by emission from highly ionized Fe. On both days, the addition of the Gaussian profile yielded a significant improvement to the fit at a confidence level of >99%, evaluated with the standard F-test. The resulting parameters and fluxes from the power-law fits are given in Table 2. XSPEC software was used to determine confidence limits on $N_H$.

## 3. RESULTS

### *3.1) The X-Ray Spectrum*

The Ginga spectra taken on 1988 May 18 and May 22/23 are shown in Figure 1 along with the best fits for a power-law model with an Fe emission line. The May 23 spectrum is of inferior quality due to higher local background during this observation. The results of all X-ray spectral fits are given in Table 2, along with the HEAO-1 flux measurements for comparison. The best fit log slope from 2 to 20 keV is $-0.80 \pm 0.02$ for 1988 May 18 and $-1.02 \pm 0.10$ for May 22/23. The Gaussian emission lines are centered at $6.19 \pm 0.28$ keV for May 18 and $6.50 \pm 0.27$ keV (90% confidence limits) in the rest frame of the source for May 22/23, and so we interpret this feature as Fe fluorescence at 6.4 keV. Many observations with Ginga show that X-ray Fe emission lines are common in Seyfert galaxies (Makino 1987, Ohashi 1988, Pounds 1989).

The spectral fits did not constrain the HI absorption column well. On May 18, the 90 % confidence interval had an upper bound of $3.58 \times 10^{21}$ cm $^{-2}$. On May 23, when the background subtraction was of poorer quality, the 90% confidence interval had an upper bound of $6.67 \times 10^{21}$ cm $^{-2}$. Ginga has poor response below 2 keV and normally cannot detect $N_H$ less than ~$10^{21.5}$ cm$^{-2}$ for objects with flux similar to H2106-099. The X-ray measurements given here are therefore not of sufficient quality to convincingly demonstrate absorption due to gas.

Comparison of the results from May 18 and May 22/23 shows that the change in the normalization and the power law slope was not significant, with deviations in the Log slope of $\leq 2\,\sigma$ (see Table 2). In conclusion, the X-ray spectrum of H2106–099 is slightly steep but not very unusual, and shows no significant evidence for spectral variability on a time scale of about five days.

The Ginga data were taken during a high background epoch, and were therefore not sensitive to small amplitude variations in flux. Large amplitude variations (i.e. two or more consecutive points with deviations of ≥50% of the mean flux) are ruled out by our data on time scales ≥ 128 s. The mean flux values from each of the two Ginga observing intervals appear consistent, with a variation of only $(-1.7 \pm 1.2) \times 10^{-12}$ ergs s$^{-1}$ cm$^{-2}$ (2-10 keV). The average Ginga flux ($6.9 \times 10^{-12}$ ergs s$^{-1}$ cm$^{-2}$, 2-10 keV) falls between the two HEAO-1 fluxes (see Table 2 and the lower panel of Fig. 2), so no significant X-ray variability is demonstrated.

### *3.2) The UV Spectrum*

#### *3.2.1) 2175 Å Absorption and Reddening Correction*

The IUE (The International Ultraviolet Explorer satellite) spectrum, taken on 1988 May 18, nearly coincident with the first Ginga observation, is shown in Figure 3. The geocoronal Ly$\alpha$ line has been replaced by a linear interpolation between points surrounding the feature. No reddening corrections have been applied to this spectrum. The strongest emission lines (see Table 3a) are typical for a Seyfert Type I. Emission from HeII is present at $\lambda$1644 Å, and is unusually strong.

The most unusual feature of the UV spectrum is a strong, broad dip between 1700 and 2300 Å. This feature is in the same location as the dust absorption feature which dominates the UV extinction function, centered near 2175 Å ($1.38 \times 10^{15}$ Hz) with a FWHM of ~480 Å (Savage and Mathis 1979). The bottom of the dip in the spectrum of H2106–099 is near 2175 Å, and such features are also present in other AGN. These features are clearly present, for example, in four UV spectra presented in Wu, Boggess, and Gull (1983), particularly in AKN 120, and in a number of spectra in Kinney et al. (1991). The features in Kinney et al. (1991), presented in an $F_\lambda$ representation, are similar to the one shown in Figure 3, but not as pronounced. In the $F_\nu$ vs $\nu$ representation, as shown in Wu, Bogess, and Gull (1983), the low-frequency side of the feature is emphasized, appearing as a precipitous drop at $\sim 1.3 \times 10^{15}$ Hz (~2300 Å). The feature in H2106–099 is obvious in both representations (but the $F_\nu$ representation is not shown here).

Initially, the near IR through IUE data were corrected for reddening within our galaxy. The extinction function used in this correction, and throughout the rest of this paper, was produced by interpolation of the function values given at discrete wavelengths in Savage and Mathis (1979). We also used a value of $4.96 \times 10^{20}$ cm$^{-2}$ for $N_{HI}$ (Stark et al. 1992) , and a conversion of $N_{HI}$ to $E(B–V)$ , the extinction difference in magnitudes between the B and V bands, of $5.2 \times 10^{21}$ cm$^{-2}$ mag$^{-1}$ (Van Steenberg & Shull 1988). The extinction along the line of sight to H2106–099 due to the gas within our galaxy is therefore given by $E(B\text{-}V) = 0.095$ mag, with an uncertainty of ~ .01 mag (following Elvis et al. 1986). Extinction is corrected by applying the following expression:

$$F_{\lambda corr} = F_\lambda \, 10^{(E(B-V) \times a_\lambda / 2.5)} \quad . \qquad (1)$$

In order to correct for galactic absorption, $F_\lambda$ is the flux of the raw spectrum at the observed wavelength λ, $E(B\text{-}V)$ is only that due to material *within our galaxy*, $E_{gal}(B–V)$ ,

$F_{\lambda corr}$ is the flux at the observed wavelength λ corrected for reddening, and $a_\lambda$ is the extinction per magnitude of *E(B–V)* at the wavelength λ (i.e. $A_\lambda = a_\lambda \times E(B\text{-}V)$).

A "close-up" view of the 2175 Å region in the spectrum of H2106–099 is given in Figure 4. The top panel, Figure 4a, is uncorrected, and the middle figure, Figure 4b, is corrected for reddening due to the material in our galaxy in the line of sight to H2106–099. Figure 4b shows that the 2175 Å feature persists, even after correction for reddening in our galaxy.

The appearance of a dip-like feature can be produced by coincidental combinations of a steep continuum and a broad emission line. The nearest broad emission feature which might contribute to this effect is the Fe II mound near 2400 Å. Examination of UV Fe lines in AGN without residual 2175Å absorption shows emission from about 2270 Å - 2650 Å (in the emitted frame; full width at zero intensity) in the very strongest, broadest cases (See Mkn 290 in Netzer et al. 1985). The Fe II emission in these objects and in the models given in WNW therefore *cannot* be responsible for the red side of the dip observed in H2106-099, which rises steeply from a wavelength of less than 2200 Å. Also, the continuum of H2106-099 cannot be responsible for the blue side of this dip. The continuum does not rise dramatically to the blue anywhere except in this region; in fact, it is rather flat all the way out to Lyα. We conclude that the residual feature (shown in Figure 4b) is due to extinction by material outside our galaxy, probably associated with the AGN or its host galaxy.

Because the residual absorption most probably occurs in the AGN rest frame, all discussion hereafter refers to the spectrum of observed flux density at the emitted wavelength. If one assumes a smooth underlying power law continuum in the UV-optical region, a good estimate of the extinction required to produce the absorption feature can be obtained by correcting the spectrum until the 2175 Å absorption feature vanishes and the continuum becomes a smooth power law.

It is difficult to judge the shape and strength of the continuum in general in the UV because of line emission and noise, so we produced a smooth model continuum by fitting points far from the 2175 Å feature. To determine the underlying continuum without any contributions from the apparent Fe II continuum, we used the results of detailed atomic modeling by WNW and Netzer et al. (1985) to define fit regions free from Fe emission: 1330-1380 Å, 1430-1460 Å, 5553 Å-5664 Å, 6750 Å-6840 Å, all in the emitted frame. The power law fits were not of good quality in the region immediately surrounding Hγ (λ 4340 Å), as the power law went above the apparent continuum in this region. When points in the regions immediately surrounding Hγ were added to the fit, regions where WNW determined some, but relatively little, Fe emission was present, the quality of the fits improved. The Fe-free regions given above plus two regions immediately surrounding Hγ, 4247-4267 Å and 4416-4460 Å (emitted frame), were used in all fits described hereafter. (Note that a narrow dip immediately to the red of CIV(λ1550 Å), shown in Figure 4, is not fit, as it is believed to be spurious. The dip does not appear in the 1987 spectrum of the object from the IUE archives, it is suspiciously close to the instrumental feature, and other lines do not show similarly related features.)

To determine the additional reddening intrinsic to the AGN or its host, the value of $E_{\mathrm{int}}(B{-}V)$ , the spectrum was de-reddened in finite steps, fitting a power law at each step, iterating until the power law intersected the continuum in an emission-free region near the 2175 Å absorption feature. In detail, the following procedures were performed on the UV-Optical spectrum from 1988 May, corrected for Milky Way reddening, then shifted to the emitted wavelength :

First, the data in the dip region from 1650-2200 Å, after emission line removal, were fitted to a Gaussian centered near 2175 Å, avoiding possible Fe II emission near 2400 Å. The Gaussian function provided a good fit to the data, however, results were not sensitive to the form of this fit, i.e. various smoothed versions of the data yielded the same final results.

Next, for each guess of $E_{int}(B–V)$, the spectrum was de-reddened by that value and a new power law was fit to the data. In this step, the spectrum is again de-reddened using expression (1), but first the data are shifted to the emitted frame where the absorption takes place.

The power-law fit intersected the lowest part of the continuum near 2175 Å for $E_{int}(B–V)$ = 0.07 mag. The log slope of this power law was –1.01. The measurement error of the data in this region, $\sigma_{2175}$, dominates the errors in this calculation, including the fit uncertainties. The value of $\sigma_{2175}$ can be estimated by the standard deviation of the residuals of the gaussian fit described above. The range of E(B-V) corresponding to +/- 1 $\sigma_{2175}$ error in the data at the intersection of the continuum and fit is 0.05-0.12 mag. The bottom panel of Figure 4 shows a "close-up" view of the 2175 Å region corrected as described above. The model power law (represented by a dotted line) intersects the actual data near 2175 Å indicating that the proper extinction correction has been applied. The figure shows a clear improvement in the smoothness of the measured continuum. The corrected optical-UV spectrum is presented in Figure 5; the fit points are indicated by small squares. The $E_{int}(B–V)$ value derived above, 0.07 mag, is of the same order as the reddening due to our galaxy, $E_{gal}(B-V)$ = 0.095 mag.

### *3.2.2) The Blue Bump*

A convenient measure of blue-bump strength relative to the continuum at 1.25 μm is given by a spectral slope between 1.25 μm and the UV continuum flux at the highest observed frequency . To improve the signal to noise ratio, we took an average of the log slopes from the 1.25μm (J band) band to three bands in the UV which exclude strong emission lines and are blue enough to avoid FeII emission (WNW); this average slope is referred to as $<\alpha_{J\text{-}UV}>$. The three UV bands are 1290–1360, 1425–1520 and 1700–1860Å in the AGN rest frame. The motivation of this average log slope is to give a measure of relative strength of the blue bump to that of the non-thermal continuum. The J band was chosen at the low frequency end because the 1.25 μm band is identified with non-thermal emission, as shown by its excellent correlation with the hard X-rays (e.g. Ward et al. 1987).

The value of $<\alpha_{J-UV}>$ for H2106–099 is –0.94 ± 0.05 after correction for reddening as given above. For an un-reddened comparison sample, we used 21 AGN from the HEAO-I catalog (Wood et al. 1984) which are known to have no significant 2175 Å absorption feature (Grossan 1992). The mean value of $\alpha_{J-UV}$ from this sample is –0.70 (Grossan 1992), and the bump of H2106–099, even after reddening correction, is weaker than all but 5 AGN in this sample. The shape of the optical-UV continuum in H2106–099 is certainly unlike that in typical strong bump objects such as 3C273. There is no rise in slope in $\nu F_\nu$ from the optical through the UV; the UV spectrum is steeper than –1 everywhere. The only "bump" observed in the optical-UV region lies in the region of the Balmer continuum, between 2000 and 4000Å. Even after correction for reddening, H2106-099 is definitely in the weak bump group of AGN.

### *3.3) Optical and Other Bands*

The optical spectrum of the object, taken with 12 Å resolution on 1988 May 23, contemporaneously with the Ginga observation, shows broad HI emission lines typical of Sy 1 nuclei, with Fe II emission, broad He I, and a number of very high excitation lines such as [FeVII] and [Fe X] (see Fig. 6). No reddening corrections have been applied to this spectrum. Optical Fe II emission is manifested as two broad bumps on either side of the Hβ-[OIII] group, centered around 4750 and 5400 Å, respectively. For this observation, Hβ/Hα = 0.27 ± 0.04 (not corrected for reddening). [NII] is not detected. Table 3b gives the emission line strengths for spectra taken in the various epochs of observation.

Emission line fluxes were simply determined via subtraction of a linearly interpolated continuum from the total flux in an emission line; no deblending was performed (for details and errors, see Grossan 1992). This approach is imperfect because of possible contamination by underlying Fe II emission (WNW), however, it is adequate for studying variability with modest sensitivity. Significant spectral variability was measured in most lines. All spectral lines appear strongest in 1982 October. Hβ and Hα vary (from maximum to minimum flux) by more than 25 % relative to the mean of the four measurements. However, the most severe spectral changes take place in He I and possibly [FeX], which changed by more than 100% (at 4-5 σ) relative to the mean between 1982 and 1988.

In the V optical band where our data is most complete, variability is observed by a factor of 2.2 (a change of 4.5 σ) for two points ~ 5 years apart, and a factor of 1.8 ( > 10 σ) within 6 weeks. Table 1 gives a summary of V band data, which is plotted in Figure 2. The X-ray and optical bands are not well sampled in the same epochs, therefore our data are not appropriate to determine the correlation between X-ray and optical flux.

Table 1 gives the results of our radio and near IR measurements. The two VLA measurements at 6 cm show ~< 25% variation (at the 4 σ level) between measurements four years apart. The two JHK measurements, separated by ~ 5 months, showed no variability.

### *3.4) The Corrected 1988 May Multi-Frequency Spectrum*

The composite spectrum of the H2106-099 appears in Figure 7, plotted as $\nu F_\nu$ vs. $\nu$. It is shown de-reddened as described in section 3.2, using $E_{int}(B–V)$= 0.07 mag for the intrinsic reddening associated with the object, and $E_{gal}(B–V)$= 0.095 mag for the local reddening due to our galaxy. The 1988 May results are represented by circular symbols. Outlined square points represent low frequency data taken before 1988 May. The IUE data are represented by a series of points averaged over roughly 250 Å bands with strong emission lines removed. In the optical bands, the emission lines contribute only a small part of the total flux in any band, so the optical photometry points have not been corrected for line emission.

Although the 1988 May data (circles) are not truly simultaneous, there are good reasons to consider them representative of the "instantaneous " spectrum of the object. The first day of Ginga observations and the IUE observations overlap. The results from the second day of Ginga observations are consistent with the flux from the first day's results, and are concurrent with optical spectrophotometry. The IRAS and JHK fluxes are not simultaneous with the higher energy data. However, variability is observed to generally decrease with wavelength, and large variations in JHK and IRAS fluxes are uncommon (Urry 1988).

The six centimeter radio emission is below the B band flux by several orders of magnitude (and is therefore not shown in fig. 7). This object is therefore radio-quiet, and any radio-linked components (e.g. Barvanis 1990) are weak. The IR spectrum is nuclear-dominated, showing no evidence for strong contamination by thermal components. The lack of a steeply falling IRAS spectrum in the $\nu F\nu$ representation (Ward et al. 1987), and the lack of strong peaks in the IR and near IR regions suggest that the contributions from cool dust in the host galaxy disk, which peaks at around 100–60 μm ( $<1.8 \times 10^{12}$ Hz ), and from warm nuclear dust, which peaks at around 5 μm ( $6 \times 10^{13}$ Hz ) (Barvanis 1990), are small. The weakness of the 1 μm dip also suggests that there is not strong emission by warm dust. Starlight, which contributes a characteristic bump between about 0.4 and 3 μ (Barvanis 1990), must also be weak in this object because the bump is not present.

## 4. DISCUSSION

### *4.1) General*

The 2-10 keV X-ray luminosity of most of the identified AGN from the HEAO-I sample ranges from $\sim 10^{40.5}$ - $\sim 10^{47.5}$ erg s$^{-1}$ ($H_0$/50 km s$^{-1}$ Mpc$^{-1}$)$^{-2}$ (Grossan 1992), so the luminosity of H2106–099, $10^{43.7}$ erg $s^{-1}$, is typical for AGN detected in this band. The object is unusual, however, because it has high excitation Fe lines, and no strong blue bump. H2106-099 is also unusually variable in the optical bands for a non-blazar AGN.

### *4.2) Blue Bump Comparisons*

In the following discussion $<\alpha_{J\text{-}UV}>$ is taken to be an ad hoc approximate indicator of blue bump "strength", for purposes of comparison (while recognizing that this is correct only if homogeneous distributions of emission temperature and contributions of non-thermal flux are assumed). If a -1 log slope non-thermal component is assumed, $<\alpha_{J\text{-}UV}> = -1$ indicates no blue bump, and larger values of $<\alpha_{J\text{-}UV}>$ indicate greater bump strength. There are several examples of AGN with apparently weaker bumps than $<\alpha_{J\text{-}UV}> = \sim-0.6$ among the HEAO-1 objects (Grossan 1992), but such objects are rare in the literature (e.g. McDowell et al. 1989). (Several works show apparently steep blue spectra, but few include actual measurements covering the full IUE band, e.g. Sanders et al. 1989, and these cannot be directly compared). The bump strength of H2106–099, $<\alpha_{J\text{-}UV}> = -0.94 \pm 0.05$, is among the weakest of those reported in the literature with full IUE coverage. The lack of apparently weak-bump non-X-ray selected objects in the literature is probably due to the expected bias against reddened objects in optical surveys and identification. As pointed out in McDowell et al.(1989), the reddening due to $E(B\text{-}V) = 0.1\text{-}0.2$ mag is enough to move the objects appearing to have the weakest bumps back into the category of average bump strength. Reddening is therefore always a concern in blue bump measurements, because the most commonly used reddening indicators are not reliable for $E(B\text{-}V) < 1$ mag (see section 4.4). The identification of objects with weak bumps (after de-reddening) is important because it allows us to test UV spectral models, such as those with inclined and/or obscured disks (or perhaps models without disks). Given the assumptions above, the nearly bumpless H2106-099 is a challenge for the simplest "dusty torus" models (e.g. Woltjer 1990), which would predict obscuration of the BLR and absorption of the central X-ray source in order to provide a weak blue bump by obscuring the accretion disk.

*4.3) Optical Spectral Variability*

Winkler (1992) shows two spectra of H2106–099 taken three years apart, and significant variability of Hβ is evident. The same paper explicitly contradicts the finding in Remillard et al. (1986) that He I is unusually strong in this object. Our measurements verify that He I was much brighter in 1982 than in any subsequent spectra. Shuder (1982) argued that He I was located very close to the nucleus, because its width is systematically larger than that of Hα and Hβ. Perhaps this line is highly variable because it is closest to the central engine, and therefore most directly affected by variations in the ionizing flux.

*4.4) Reddening Correction*

There are fundamental problems with the use of X-ray $N_H$ values and the Balmer decrement as reddening indicators in the correction of UV AGN spectra. High $N_H$ values within our galaxy imply reddening, but such values could be accompanied by no reddening in an AGN; there are large regions where the temperature is high enough to evaporate any dust. Next, consider the Balmer decrement. The intrinsic spread in the value of the decrement causes large uncertainties in the inferred reddening. In addition, line emission may not suffer the same extinction as the continuum components, so the decrement could measure the wrong reddening for the continuum. Finally, consider the measurement precision required to use the decrement for reddening correction of UV spectra. To make a 2 σ measurement of the intrinsic reddening for H2106–099, Hα and Hβ must be measured to an accuracy of better than 3 %. The contribution of Fe II in this measurement, by itself, is probably not known to this accuracy, but noise and measurement systematic errors must be considered as well, and much larger errors are the norm. An error of 10 % in the line fluxes corresponds to an error in $E(B\text{-}V)$ of ~0.13 mag in this object, which would change the flux near Ly$\alpha$ by a factor of 2.7. The Balmer decrement is therefore a hopelessly insensitive indicator for the purpose of UV reddening correction.

To use the method of de-reddening presented here, we assumed that the reddening function given is applicable to the material along the line of sight in both the Milky Way and at the AGN and host. Significantly different extinction functions have been reported for the LMC (Nandy et. al. 1981) and the SMC (Prevot et al. 1984), with weak 2175 Å features. These systems, classified as irregulars, are *very* different from our own galaxy. It is plausible that the environment in these galaxies would result in different dust grain properties, and therefore a different extinction function. The morphology of Seyfert galaxies is predominantly spiral (e.g. Kirhakos & Steiner 1991), like our own galaxy, which suggests that the galactic extinction curve is appropriate. Finally, the presence of the prominent 2175 Å feature eliminates the possibility of SMC-like extinction in H2106–099, and supports the assumption of Milky-Way like extinction.

There is a clear lesson from this reddening investigation of H2106-099: before *any* comparisons of the UV continuum to that in other bands, such as in blue bump strength measurements, the 2175Å region had better be checked for evidence of reddening or it may be missed. This object has a low Balmer decrement, the 21 cm $N_{HI}$ measurement indicates typical reddening for the object's galactic latitude, and an X-ray observation does not show a high absorption column; yet the presence of the residual 2175 Å feature indicates significant reddening due to material outside of our galaxy.

### *4.5) Continuum Spectral Components*

Barvanis (1990) suggests that for radio quiet quasars (and presumably AGN), the region from 1 $\mu$ through the UV can be fit, after removal of dust and starlight contributions, by a single power law with a log slope of $\alpha_{IU} = -0.2$, the index expected for free-free emission. Our data extend much farther to the UV than that in Barvanis (1990). Not only is our J-UV log slope much steeper than the index Barvanis suggests, but we find that fits to a power law, with almost any subset of our data anywhere between 1 μm and the UV (excluding the 2000-4000 Å Balmer continuum region) are always much steeper than –0.2. This result suggests that free-free emission is not important in this object in this region of the spectrum. Barvanis treats one object with a very steep slope, Mkn 231, and he attributes the slope to internal reddening. Even at the high end of acceptable reddening values, an energy exponent of –0.2 would still not be attained between 1 μm and the UV in H2106-099.

After corrections for reddening were applied, we showed that the blue bump in H2106-099 is quite weak. It is therefore possible that the apparent UV continuum in this object is a non-thermal component normally masked by the blue bump in other objects with stronger bumps. The data presented here are consistent with the model of an underlying non-thermal near IR-X-ray power law with a slope of –1.0 (Ward et al. 1987). If all AGN known with *apparent* slopes steeper than –1.0 were re-examined and corrected for reddening via 2175 Å observations, then one could form a stringent test of this model. If any of the corrected spectra were significantly steeper than –1, then either there is no underlying power law or it is steeper than a –1 log slope. If no objects with slopes steeper than –1 remained, the model would be empirically verified.

## 5. SUMMARY

1. H2106-099 shows significant spectroscopic and photometric variability. The optical V band was observed to change by a factor of 1.8 (>10 σ) in six weeks. The Balmer line flux was observed to change by > 25% (>30% in equivalent width), and the He I flux was observed to change by greater than 100% in both equivalent width and flux over ~ six year time scales.

2. We used an AGN spectrum de-reddening technique based on observations of the UV region including 2175 Å in the emitted frame, and an assumed power-law continuum. We found that H2106-099 required an additional reddening correction of 0.07 mag above that due to the material in our galaxy.  Neither an anomalous Balmer decrement nor any other indication of reddening was found except for the absorption dip near 2175 Å.  These observations suggest that reliable measures of the blue bump strength, or any other measurements dependent on absolute UV line or continuum flux (which are strongly sensitive to reddening), cannot be made without first checking an observation of the 2175 Å region in the emitted frame for an absorption feature, and then correcting for the reddening indicated.

3. No strong blue bump is present in this object, whether measured by the value of $<\alpha_{J\text{-}UV}>$, or by the qualitative appearance of the blue bump region.  The $<\alpha_{J\text{-}UV}>$ for this object is –0.94 ± 0.05 (from our best reddening correction value, $E_{int}(B\text{-}V)$ = 0.07 mag), which indicates a weak blue bump compared to that of other AGN observed over the same bands, and which is far below the $\alpha_{J\text{-}UV}$ values ascribed to "strong bump objects" such as 3C 273.  Because the weak blue bump in H2106–099 has been shown to be *intrinsic*, not due to internal reddening, weak blue bump objects should now be considered a bona fide subclass of AGN.

## ACKNOWLEDGMENTS

The authors wish to thank the Staff of the Michigan-Dartmouth-MIT Observatory, the staff and students of the Institute for Space and Astronomical Sciences (ISAS), of Sagamihara, Japan, for providing invaluable assistance to the foreign visiting astronomers, and to the Ginga, HEAO, and IUE teams for taking on the difficulties of international collaboration, which made this paper possible. Thanks also to Rees Williams for many late-night discussions regarding the inner workings of Ginga data analysis.

**TABLES**

**TABLE 1**

Journal of Observations

| Date (UT) | Band / Instrument[a] | Flux in Given Band |
|---|---|---|
| | **X-Ray** | $F_X$ (μJy @5 keV) |
| 1977 November 5-11 | HEAO LASS scan 1 | 1.2 ± 0.2 |
| 1978 May 3-9 | HEAO LASS scan 2 | 0.39 ± 0.2 |
| 1988 May 18.03 — 18.64[b] | Ginga Satellite | 0.77 ± 0.03 |
| 1988 May 22.93 — 23.47 | Ginga Satellite | 0.60 ± 0.12 |
| | **Ultra Violet** | Flux @~1330Å (mJy) |
| 1987.5 | IUE | 0.82 ± 0.05[c] |
| 1988 May 18[b] | IUE | 0.58 ± 0.04[c] |
| | **Optical** | V band (mag) |
| 1982 Oct 18 | Photometry, CTIO[d] | 14.32 ± 0.03 |
| 1982 Oct 25 | I.D. Spectrum, MDM[d,e] (15Å) | … |
| 1987 May 18 | Optical Photometry[g] | 14.66 ± 0.05 |
| 1987 July 4 | Optical Photometry | 14.57 ± 0.03 |
| 1988 May 21, 22, 23[f] | Optical Spectra, MDM[e] (12 Å) | 14.62 ± 0.1 |
| 1988 May 27, Jun 1 | Optical Photometry, MDM[e] | 14.67 ± 0.02 |
| 1989 May 14-21 | Optical Photometry and Spectra, MDM[e] (6 Å) | 14.41 ± 0.01 |
| 1989 Sep 29 | Optical Photometry, MDM[e] | 14.23 ± 0.01 |
| | **Near Infra-Red** | J Band (mags) |
| 1987 May 22 | Near IR photometry (J band )[g] | 12.54 ± 0.05 |
| 1987 Oct 27 | Near IR photometry (J band )[g] | 12.53 ± 0.02 |
| | | |
| | **INFRA-RED (IRAS)** | Flux (mJy) |
| 1983 | 12, 25, 60, 100 μm | 130±26, 350±45, 270±21, <1980 (3 σ upper limit) |
| | | |
| | **Radio** (frequency) | Flux (mJy) |
| 1984 | VLA[d] (1465MHz) | 3.6±0.3 |
| 1987 June 24 | Parkes[g] (8.40 GHz) | 5.0±2.0 |
| 1988 May 3,8 | VLA[g] (1465, 4885 MHz) | 4.76±0.01, 1.34±0.02 |

[a] Numbers in parenthesis in this column give the resolution, in Å, for optical spectroscopy measurements.

[b] Nearly simultaneous observations.

[c] Average flux from 1290 - 1360Å. The systematic error was assumed to be 5 % for these measurements.

[d] Remillard et al. (1986) .

[e] "MDM" refers to data taken at the Michigan-Dartmouth-MIT Observatory on Kitt Peak in Arizona.

[f] Observations simultaneous with the *Ginga* observation, derived from spectrophotometry using 10” apertures (all other results are photometric).

[g] Brissenden (1989).

Table 2

X-Ray Spectra Fit Parameters[a]

| Parameter | 1988 May 18 - Ginga | 1988 May 22/23 - Ginga |
|---|---|---|
| Coefficient 'A' ($(\mathrm{keV})^{-\alpha+1}$ $\mathrm{keV}^{-1}$ $(4000\ \mathrm{cm}^2)^{-1}$ $\mathrm{s}^{-1}$)[b] | 16.50 ± 0.56 | 18.1 ± 2.9 |
| $\alpha$[b] | $-0.80 \pm 0.02$ | $-1.02 \pm 0.10$ |
| Gaussian Profile | | |
| Center Energy(keV) | 6.03 ± 0.27 | 6.33 ± 0.26 |
| Equivalent Width (keV) | 0.115 ± 0.049 | 0.133 ± 0.057 |
| $N_H$($\mathrm{cm}^{-2}$) | | |
| Lower Limit, 90%[c] Conf. | $1.00\times10^{18}$ | $7.84\times10^{20}$ |
| Upper Limit, 90%[c] Conf. | $3.58\times10^{21}$ | $6.67\times10^{21}$ |
| Best Fit | $1.33\times10^{21}$ | $3.70\times10^{21}$ |
| $\chi^2_\nu$[d] | 1.04 | 1.19 |
| Flux at 5.2 keV[e] ($\mu$Jy) | 0.77 ± 0.03 | 0.60 ± 0.12 |

[a] The given parameter uncertainties are one-dimensional 90 % confidence limits unless otherwise noted.

[b]A and $\alpha$ are the parameters of the power law component of the model spectrum defined as follows: $N(E) = A\ E^{\alpha-1}$, where E is the photon energy in keV, and N(E) is the photon flux in the LAC in units of $(4000\ \mathrm{cm}^2)^{-1}$ $\mathrm{s}^{-1}$ $\mathrm{keV}^{-1}$, plotted in figures 1a and 1b. Note that $\alpha$ is then log slope of the *energy flux*, consistent with the rest of the text.

[c]The given limit is the projection of the 90% confidence contour in multi-dimensional $\chi^2$ space onto the one dimensional $N_H$ axis.

[d]Reduced $\chi^2$ for 17 d.o.f.

[e]These values may be compared to the HEAO-1 LASS measurements which imply values of 1.2 ± 0.2 μJy during the 1977 scan and 0.39 ± 0.2 μJy during the 1978 scan (Remillard et al. 1986).

Table 3a

UV Line Fluxes

| Emission Line | Flux [a] ($10^{-12}$ ergs $cm^{-2}$ $s^{-1}$) | Line Flux / Hβ Flux[a,b] |
|---|---|---|
| Lyα[c] | [2.33] | [6.74] |
| | 10.9 | 18.2 |
| C IV | [0.79] | [2.30] |
| | 2.92 | 4.85 |
| Si IV | [0.18] | [0.52] |
| | 0.68 | 1.13 |
| Mg II | [0.30] | [0.88] |
| | 0.75 | 1.24 |
| 2500Å Fe II mound[d] | [0.69] | [1.43] |
| | 2.62 | 4.81 |
| Fe II UV total[e] | [1.26] | [2.62] |
| | 6.94 | 8.59 |

Notes:

[a]The upper values, given in square brackets, were measured from the IUE uncorrected spectrum; the lower values were measured from the IUE spectrum de-reddened by $E_{gal}(B\text{-}V)$ = 0.095 mag and $E_{int}(B\text{-}V)$ = 0.07 mag., the extinction determined in the text. Note that the Fe line fluxes will not scale by the ratio of reddening corrections, because their measurement is dependent on power law fits to corrected data.

[b]For standardization purposes, we present fluxes normalized by the raw absolute flux of Hβ measured to be $4.82 \times 10^{-13}$ ergs $cm^{-2}$ $s^{-1}$ on 1988 May 23, nearly simultaneously with this IUE observation.

[c]Because of the overlap of the geocoronal line and Lyα, the high frequency wing of Lyα was interpolated linearly through this feature.

[d] The "2500Å Mound" is defined as the blended Fe II emission in the region from 2400-2700Å.

[e] "Fe II UV total" is defined as the region from 2000-3000Å, the same as is defined in WNW. Note, however, that the contribution of the Balmer continuum is not removed here, as it is in WNW.

**TABLE 3B**

# FIGURES

## FIGURE CAPTIONS

Figure 1. The Ginga X-ray spectra from 1988 May 18 and 22/23 are shown here along with the best fit model power law with absorption and a Gaussian Fe emission line (the model functions are represented by solid lines). Residuals are shown beneath each plot. Note that the May 22/23 spectrum has a poorer fit at high energies due to higher background during this observation. The photon spectra are shown in the detected frame.

Figure 2. X-ray and optical fluxes are presented on the same time axes, for two relevant time segments. Significant variability is evident only in the V band. No reddening correction has been applied. V band data from Winkler (1992) are shown as open squares; all other data is described in the text. See Table 1 for more details.

Figure 3. The IUE SWP and LWP (short and long wavelength spectrometers) spectra are presented here together (detected frame). The Lyα geocoronal line has been removed. No reddening corrections have been applied. The feature marked “A” is instrumental.

Figure 4 a-c. The 2175 Å absorption feature is shown in this figure de-reddened by $E_{gal}(B–V)$ = 0.0 mag in the top panel (a), by $E_{gal}(B–V)$ =0.095 mag, the extinction for material within our galaxy, in the middle panel (b), and by $E_{gal}(B–V)$ = 0.095 mag plus an additional $E_{int}(B–V)$ = 0.07 mag in the bottom panel (c). All wavelengths are in the rest frame of the AGN. See text for details.

Figure 5. The composite IUE and optical spectrum from May 1988 is shown, corrected for a total reddening of $E_{gal}(B–V)$ = 0.095 mag plus an additional $E_{int}(B–V)$ = 0.07 mag (see text for details), presented as observed flux in the emitted frame. A power law is also shown, fit to those points represented by outlined squares. The power law has a log slope of –1.0.

Figure 6. This figure shows the optical spectrum (detected frame) of H2106–099 on 1988 May 23 (12 Å resolution), not corrected for reddening.

Figure 7. The H2106–099 Multi-Frequency spectral energy distribution is given dereddened by $E_{gal}(B–V)$ = 0.095 mag plus an additional $E_{int}(B-V)$ = 0.07 mag. The figure shows $\nu \times F\nu$ vs $\nu$ on Log-Log axes. The observed flux is plotted at the emitted frequency. All data taken during 1988 May are shown as circles. Those data taken before 1988 May are shown as outlined squares (see Table 1 for a log of the observations). The power law fit to the 1988 May 18 X-ray spectrum is represented by a dotted line, and the 2-10 keV integrated flux, assuming the spectral parameters given in Table 2, is represented as an open circle at 5.2 keV. The IUE points are the averages of the following bands: 3030-3100, 2880-3030, 2570-2720, 2150-2325, 1990-2050, 1700-1860, 1425-1520, 1290-1360 and 1130-1170 Å. See text for full details.

Table 3b

Optical Line Flux Measurements[a]

| Date | Hγ | FeIIblue +HeI | FeIIvis | Hβ[b] | [OIII] 5007Å | [FeVII] 5721 Å | HeI | [FeVII] 6085 Å | [FeX] 6374 Å | Hα |
|---|---|---|---|---|---|---|---|---|---|---|
| 1982 Oct | 41.3 | 52.4 | 29.9 | 100.0 | 37.2 | 5.73 | 52.3 | 6.41 | 11.6 | 555 |
| | 4.31 | 3.45 | 2.40 | 2.59 | 1.74 | 2.60 | 4.65 | 3.30 | 2.50 | 5.75 |
| 1988 May 21 | 38.1 | 35.8 | 27.9 | 114 | 42.6 | 3.05 | 20.4 | 6.33 | 2.80 | 429 |
| | 2.82 | 2.88 | 1.14 | 1.99 | 1.43 | 0.62 | 1.04 | 0.31 | 0.46 | 0.95 |
| 1988 May 23 | 48.9 | 31.9 | 43.6 | 133 | 47.4 | 4.35 | 21.0 | 6.26 | 2.90 | 492 |
| | 2.34 | 3.42 | 1.11 | 2.40 | 1.72 | 0.50 | 0.91 | 0.52 | 0.40 | 1.36 |
| 1989 May | 53.8 | 44.2 | 33.2 | 122 | 41.9 | 5.04 | 21.8 | 5.20 | 1.39 | 427 |
| | 1.52 | 1.45 | 0.63 | 1.02 | 0.70 | 0.36 | 0.67 | 0.33 | 0.44 | 1.15 |
| Variation/Mean[c] | 0.34 | 0.50 | 0.47 | 0.28 | 0.24 | 0.59 | 1.11 | 0.20 | 2.18 | 0.27 |
| σ | 0.07 | 0.12 | 0.05 | 0.03 | 0.06 | 0.59 | 0.17 | 0.55 | 0.54 | 0.01 |

Notes:

[a] For the purpose of standardized comparisons, no reddening corrections have been applied to these measurements. Relative line strengths are given with Hβ in 1982 October = 100. The absolute flux of this line was measured to be $3.41 \times 10^{-13}$ erg $s^{-1}$ $cm^{-2}$ at that time. Uncertainties are given underneath each flux value. See text for definition of uncertainties.

[b] The Hβ measurements have not been corrected for contributions from FeII or any other weak line emission in the region, however, the contribution of [OIII] has been removed.

[c] Variation/Mean is defined to be the maximum difference between any two flux measurements divided by the mean.